\documentclass[twocolumn,pra,amsmath,amssymb,superscriptaddress,showpacs]{revtex4-1}

\usepackage{amsmath, amsthm, amssymb}
\usepackage{graphicx}
\usepackage{dcolumn}
\usepackage{bm}
\usepackage{color}

\def\textbf#1{{\bf #1}}
\def\be{\begin{equation}}
\def\ee{\end{equation}}
\def\ben{\begin{eqnarray}}
\def\een{\end{eqnarray}}
\def\eea{\end{array}}
\def\bea{\begin{array}}

\newcommand{\bei}{\begin{itemize}}
\newcommand{\eei}{\end{itemize}}

\def\trace{\mbox{Tr}}
\begin{document}

\newcommand{\eg}{{\it{e.g.~}}}
\newcommand{\ie}{{\it{i.e.~}}}
\newcommand{\etal}{{\it{et al.}}}
\newcommand{\daniel}[1]{{\color{red} #1}}
\newcommand{\leandro}[1]{{\color{blue} #1}}
\newcommand{\marco}[1]{{\color{green} #1}}


\newcommand{\ci}[2]{{I(#1\rangle #2)}}
\newcommand{\disc}[2]{{D(#1 | #2)}}
\newcommand{\discinf}[2]{{D^\infty(#1 | #2)}}
\newcommand{\dc}[2]{\chi_{\text{DC}}(#1\rangle #2)}
\newcommand{\Ddc}[2]{\Delta_{\text{DC}}(#1\rangle#2)}
\newcommand{\tec}[2]{\Gamma(#1\rangle#2)}
\newcommand{\tecinf}[2]{\Gamma^\infty(#1\rangle#2)}
\newcommand{\eof}[2]{E_F(#1:#2)}


\author{D. Cavalcanti}
\affiliation{Centre for Quantum Technologies, National University of Singapore, 2 Science Drive 3, Singapore 117542}

\author{L. Aolita}
\affiliation{ICFO-Institut de Ci\`{e}ncies Fot\`{o}niques, Mediterranean
Technology Park, 08860 Castelldefels (Barcelona), Spain} 

\author{S. Boixo}
\affiliation{Institute for Quantum Information, California Institute of Technology, Pasadena, CA 91125, USA}

\author{K. Modi}
\affiliation{Centre for Quantum Technologies, National University of Singapore, 2 Science Drive 3, Singapore 117542}

\author{M. Piani}
\affiliation{Institute for Quantum Computing $\&$ Department of Physics and Astronomy, 
University of Waterloo, Waterloo, Ontario, Canada, N2L 3G1}

\author{A. Winter}
\affiliation{Centre for Quantum Technologies, National University of Singapore, 2 Science Drive 3, Singapore 117542}
\affiliation{Department of Mathematics, University of Bristol, Bristol BS8 1TW, U.K.}

\title{Operational interpretations of quantum discord}
\begin{abstract}
Quantum discord quantifies non-classical correlations going beyond the standard classification of quantum states into entangled and unentangled ones. Although it has received considerable attention, it still lacks any precise interpretation in terms of some protocol in which quantum features are relevant. Here we give quantum discord its first information-theoretic operational meaning in terms of entanglement consumption in an \emph{extended quantum state merging} protocol. We further relate the asymmetry of quantum discord with the performance imbalance in quantum state merging and dense coding.
\end{abstract}
\maketitle

\section{Introduction}
The study of quantum correlations has mostly been focused on entanglement \cite{horod}. This is because entanglement has been identified as a key ingredient in quantum information processing, allowing to perform a number of tasks that are either impossible to realize or less efficient with only classical resources at disposal. However, entanglement does not account for all the non-classical properties of quantum correlations. Zurek \cite{zurek2000}  (see also \cite{henderson, ollivier}) identified \emph{quantum discord} (QD)  as a feature of quantum correlations that encapsulates entanglement but goes beyond it as it is also present even in separable states. Over the past decade, QD has been the focus of several theoretical and experimental studies addressing its formal characterization \cite{piani, *CesarEtal07, *modietal,*paris10,*Adesso10,*dvb10, ferraro}, its behavior under dynamical processes \cite{ferraro,werlangPRA, *Maziero09, *mazzola, *fanchini,*jsxu}, and its connection  with quantum computation \cite{dattashaji,*Lanyon} and quantum phase transitions \cite{sarandy,*werlangQPT}. 


QD was initially introduced in the context of the analysis of quantum measurements \cite{ollivier} and afterwards interpretations in terms of the difference in performance of quantum and classical Maxwell demons were given \cite{ZurekDemon,*Terno}. Nevertheless, a large part of the quantum information community has always been skeptical towards QD as an information-theoretic quantiÞer. This is because QD has not a clear operational interpretation in this context. That is, we lack an information-theoretic task for which the QD provides a quantitative measure about the performance in the task. Thus, without this kind of operational interpretation, QD is very often considered simply a "quantumness parameter".
 
\par In this Letter we give quantum discord its long sought operational interpretation. We relate QD to state merging (SM) \cite{merging}, a well known task in quantum information. In SM a tripartite pure state is considered, i.e., Alice ($A$), Bob ($B$), and Charlie ($C$) share (many copies of) a pure state $\psi_{ABC}$. The goal in the task is that $A$ transfers her part of the state to $B$, $\psi_{ABC} \rightarrow \psi_{B'BC}$ (see Fig. \ref{triangle}), by using classical communication and shared entanglement. Here we show that the minimal total entanglement consumed
in a process we call ``extended state merging'' (ESM) from $A$ to $B$ is exactly equal to the QD between $B$ and $C$ (with measurements on $C$). We further unravel a connection between  QD to a well-known protocol in quantum information processing: \emph{dense coding} (DC) \cite{bennett92}. DC is a task that uses pre-established quantum correlations to send classical messages more efficiently than by classical mean.

We focus on the finite-dimensional case with the three parties $A$, $B$, and $C$ sharing a pure state $\psi_{ABC}$. All bipartite and single-party states are obtained by taking the appropriate partial traces of $\psi_{ABC}$. The \emph{quantum (von Neumann) entropy} of a state $\rho$ is defined as $S(\rho) =-\trace \rho \log_2\rho$. It is the generalization to the quantum domain of the \emph{classical (Shannon) entropy} of a probability distribution $\{p_i\}$ given by $H(\{p_i\})=-\sum_i p_i \log_2 p_i$. 
We write $S(X)$ to denote the entropy of the reduced state $\rho_X$.    
Similarly, we write $H(a)$ to denote the Shannon entropy of a classical random variable $a$ distributed according to some probability distribution $\{p^a_i\}$. The latter may be the marginal probability distribution $p^a_i=\sum_jp^{ab}_{ij}$ of a bivariate (in general, multivariate) probability distribution $\{p^{ab}_{ij}\}$ of two classical random variables $a$ and $b$.

\section{Conditional entropy and coherent information}
For a bipartite system $AB$, the quantum (von Neumann) conditional entropy is defined as $S(A|B):=S(AB)-S(B)$ \cite{nielsen}. It is the quantum version of the classical (Shannon) conditional entropy $H(a|b):= H(a,b)-H(b)$. Note that both are asymmetric quantities.  $H(a|b)$ measures how much uncertainty is left---on average---about
the value of $a$ given the value of $b$. It can be written as
\be
\label{eq:classical_cond_entropy}
H(a|b)=\sum_jp^b_jH(a|b=j),
\ee
where $H(a|b=j)$ is the entropy of the conditional probability distribution $p^a_{i|b=j} :=p^{ab}_{ij}/p^{b}_j$.  It has a clear  operational interpretation as the amount of classical information  that  $A$ has to give---on average---to $B$, who knows the value of $b$, so that the latter gains full knowledge also of the value of $a$~\cite{slepianwolf}. Given this interpretation for $H(a|b)$, it is always non-negative. 

However, the situation changes drastically for  quantum states, because $S(A|B)$ can take negative values, e.g. for pure entangled states. This fact
was, for a long time, an obstacle to an operational interpretation of $S(A|B)$. On the other hand, its opposite was identified as an important quantity in the context of quantum information, and was even given a name of its own: {\it coherent information} $\ci{A}{B}:=-S(A|B)$. Coherent information was originally introduced to measure the amount of quantum information conveyable by a quantum channel \cite{Schumacher}; given that it is always non-positive in the classical case, one may say that it is a purely quantum quantity.

\section{Quantum discord}One remedy to negative quantum conditional entropy is to generalize the classical conditional entropy to quantum using Eq. \eqref{eq:classical_cond_entropy}, as was done in \cite{henderson, ollivier} by defining $S(A|B_c) := \min_{\{N_{j}\}} \sum_j p^B_j S(A|B=j)$, where the minimization is over generalized measurements $\{N_{j}\}$ \footnote{In \cite{ollivier} the generalized measurement were actually restricted to complete von Neumann measurements.}, with $N_j\geq0$ for all $j$ and $\sum_j N_j=\openone_B$. We also have $S(A|B=j) = S(\rho_{A|j})$, where $\rho_{A|j} = \trace_B (\openone_A\otimes N_B^j\rho_{AB})/p^B_j$ with $p^B_j=\trace(\openone_A\otimes N^j_B\rho_{AB})$. $S(A|B_c)$ is always positive and can also be thought of as a measure of the uncertainty  left  on average about $A$ given that  $B$ has been measured. For classical systems both $S(A|B)$ and $S(A|B_c)$ coincide with the classical conditional entropy, but in general $S(A|B_c)$ is strictly larger than $S(A|B)$. The difference in these two quantity is indeed the definition of the \emph{quantum discord with measurements on $B$} \cite{ollivier}
\be
\label{disco}
\disc{A}{B}:=S(A|B_c) - S(A|B).
\ee
QD can be seen as the gap between the standard measure for total correlations present in a quantum state $\rho_{AB}$, given by \emph{quantum mutual information} $I(A:B):=S(A)-S(A|B)$ \cite{groisman}, and the \emph{Henderson-Vedral measure of classical correlations} $I(A:B_c) := S(A) -S(A|B_c)$~\cite{henderson}. As $\disc{A}{B}=I(A:B)-I(A:B_c)$, the QD can be considered a (asymmetric) quantifier of non-classical correlations present in a quantum state. We will refer to $\disc{X}{Y}$ as to the ``discord of $XY$ measured by $Y$''.

\section{State merging and entanglement consumption}
\begin{figure}[t]
	\resizebox{8.0 cm}{4.38 cm}
	{\includegraphics{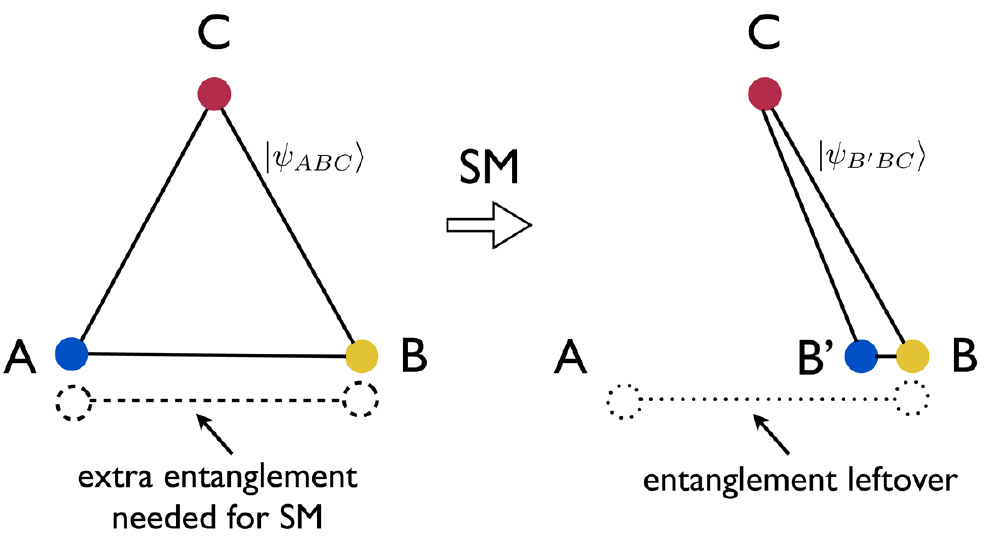}}
	\caption{\label{triangle}(Color online). Starting from a tripartite state $\psi_{ABC}$, the goal of SM is to transfer Alice's (A) part of the state to Bob (B), possibly using some extra entanglement or having some entanglement leftover. The total entanglement consumption in this process is equal to $\disc{A}{C}$ (see Eq.\eqref{eq:opmeaning}).}
\end{figure}

A fully convincing operational interpretation of quantum conditional entropy and coherent information was given with the introduction of the task of \emph{quantum state merging} (SM)~\cite{merging}. SM, say from $A$ to $B$, is a process by which $A$ and $B$ transfer $A$'s part of the state to $B$ maintaining the coherence with the reference $C$.  $A$ and $B$ both know the state they share, and they can apply arbitrary local operations coordinated by classical communication (LOCC). By acting on $n$ copies of $\psi_{ABC}$, their goal is to end up with a state close to $\psi_{B'BC}^{\otimes n}$, such that the subsystem $B'$ is in Bob's hands and plays in the new state exactly the same role as $A$ played in the old one. Errors are allowed, but they must vanish in the limit $n\rightarrow\infty$. To achieve their goal, $A$ and $B$ are allowed to use extra, pre-established two-qubit maximally-entangled pairs (ebits), but these constitute a valuable resource they must pay for. It turns out that the value of $S(A|B)$ quantifies exactly the optimal amount---per copy of the state---of ebits spent in the process. A positive value means that entanglement must be consumed, while a negative amount means not only that no extra entanglement is needed, but also that $A$ and $B$ retain $-S(A|B)=\ci{A}{B}$ ebits per copy merged. See Fig. \ref{triangle} for an illustration of SM.

A useful way to think of the role played by the conditional entropy in SM is to imagine a hypothetic entanglement bank in which $A$ and $B$ possess a joint account: the entanglement balance after merging---in ebits, per copy merged---is given precisely by $-S(A|B)$.  When $S(A|B) \geq 0$, $A$ and $B$ have to withdraw $S(A|B)$ from their account to perform SM. On the other hand, when $S(A|B) < 0$ then the process can be completed without any withdrawing. Moreover, after merging they end up sharing $\ci{A}{B} = -S(A|B)$ extra ebits of entanglement, which they deposit in their account for future use. 

At the end of this process, the only correlations
between A and B are those present in the bank account.
In particular, there is no additional entanglement left
between A and B. Given this, the bank-account picture
suggests to consider a more comprehensive balance, that
takes into account also the entanglement ``lost" in the
process. Indeed, coherent information is positive only if
the state is entangled, and while A and B may end up
with ``leftover" Bell pairs after SM, they do not share
anymore the starting entangled states.
Thus, it is useful and sensible to define the \emph{total entanglement consumption} as
\be
\label{cost1}
\tec{A}{B}:=
 \eof{A}{B} + S(A|B),
\ee
where $\eof{A}{B}:=\min_{\{p_i,\psi^{AB}_i\}} \sum_i p_i S\big(\trace_{A} (\psi_i^{AB}) \big)$ is the \emph{entanglement of formation} (EoF) of $\rho_{AB}$, with the minimum taken over pure-state ensembles $\{p_i,\psi^{AB}_i\}$ for $\rho_{AB}$~\cite{bennett}. EoF quantifies the minimum amount of pure-state entanglement that $A$ and $B$ need to consume to create $\rho_{AB}$ by LOCC with strategies where each pure-state member of the ensemble forming $\rho_{AB}$ is prepared independently. Thus, $\Gamma$ quantifies the total entanglement consumed in SM, by taking into account the amount of entanglement $A$ and $B$ would have needed to prepare $\rho_{AB}$ by LOCC---and ``lost'' during SM---plus the amount of entanglement used by the process of SM itself. In order to give a more precise operational interpretation, we consider a two-step process.
In the first stage, Alice and Bob prepare the state $\rho_{AB}$. To this aim, they
have to share classical information, and potentially use
some other local ancillas. We demand that, in order to
end up sharing $\rho_{AB}$ and not some larger state, after preparing
the state and before the merging, they remove all ancillas.
Then Eq. (3) indeed characterizes the entanglement cost
of a two-stage process that we call extended state merg-
ing (ESM): (i) state preparation through the (possibly
non-optimal -- see section Regularization below) protocol
described before and (ii) merging.

%

\section{Operational interpretations of quantum discord}

\subsection{Quantum Discord and Extended State Merging}
Now we are in the position to give QD an operational interpretation. In Appendix 1 we prove the following:
\be
\label{eq:opmeaning}
\disc{A}{C}=\tec{A}{B}.
\ee
This equation says that QD between $C$ and $A$ with measurements on $C$ is equal to the total entanglement consumption in ESM from $A$ to $B$. To the best of our knowledge, this yields the first information-theoretic scenario where the value of QD provides concrete quantitative information about a task's performance or cost.

\subsection{Asymmetry of quantum discord}One immediate exercise of the last equation is to give meaning to the asymmetry of QD, that is, the fact that in general $D(A|C)\neq D(C|A)$.  Thanks to Eq. \eqref{eq:opmeaning} we can interpret the asymmetry of discord as the differences in the cost of ESM for $A$ versus $C$ to send their parts of the state to $B$, \ie:
\be\label{eq:discasym}
D(A|C)-D(C|A)=\tec{A}{B}-\tec{C}{B}.
\ee

\subsection{Quantum Discord and Dense Coding}
Coherent information also describes the usefulness of a quantum state $\rho_{AB}$ as a resource for dense coding (DC) \cite{bennett92}. DC---say from a sender $A$ to a receiver $B$, initially sharing  $\rho_{AB}$---is a procedure by which  $A$ is able, by sending her subsystem to $B$, to transmit more classical information than she could if the system was classical; i.e., the maximal rate of classical information transmission per copy of $\rho_{AB}$ used can be larger. If $A$'s encoding is done by unitary rotations, the correction to the classical capacity that she could achieve by sending a classical system with dimension equal to that of her subsystem, $d_A$, is exactly the coherent information $\ci{A}{B}$~\cite{DC,winterdense,Bruss,HoroPiani}. In the most general DC scenario  \cite{DC,winterdense,HoroPiani}, $A$ encodes her message by means of general quantum operations $\Lambda_A:M_{d_A}\rightarrow M_{d'_A}$, where $d_A$ is the dimension of the original subsystem in the hands of $A$, while $d'_A$ is the dimension of the subsystem sent to $B$, and $M_d$ denote the set of $d\times d$ complex matrices. If the encoding is applied at the level of single copies of the shared state $\rho_{AB}$, the DC single-copy capacity can be achieved by a unitary encoding after a pre-processing operation whose aim is exactly that of increasing coherent information. More precisely the capacity is equal to $\dc{A}{B}:=\log_2 d'_A+\max_{\Lambda_A}\ci{A'}{B}$, where the maximization is over all quantum operations with output dimension $d'_A$ and $\ci{A'}{B}$ is the coherent information of $(\Lambda_{A}\otimes\openone_B)[\rho_{AB}]$. This capacity depends on the output dimension $d'_A$, but,  given that $\log_2d'_A$ can be considered as a classical contribution, one can focus on the \emph{quantum advantage of DC} 
\be
\label{eq:advantage}
\Ddc{A}{B}:=\max_{\Lambda_A}\ci{A'}{B}.
\ee
The maximization above has no restriction on the output dimension, which can anyway be taken to be less or equal to $d_A^2$~\cite{HoroPiani}. 
The maximization over $\Lambda_A$ ensures that the coherent information of the pre-processed state is non-negative. 

In Appendix 2 we prove the following connection between QD and DC: 
\be\label{eq:Discoop1}
\disc{A}{C}-\disc{B}{C}=\Ddc{C}{A}-\Ddc{C}{B}.
\ee
Note that, if $C$ sends subsystems with the same dimension to $A$ and $B$ (in particular a dimension large enough to achieve the quantum advantage of DC with both receivers) this difference can be written as
$\disc{A}{C}-\disc{B}{C}=\dc{C}{A}-\dc{C}{B}$, i.e., in terms of the DC capacity itself.


Eq.~\eqref{eq:Discoop1}  gives an operational meaning in terms of performance to the differences in QD: 
the difference in the QD of $AC$ and $BC$, both measured by $C$, is the same as the difference in the DC capacity from $C$  to either $A$ or $B$. 
The same difference in QD can be related to the coherent information, as can be seen using Eq.~\eqref{eq:opmeaning} twice: $\disc{A}{C}-\disc{B}{C}=\ci{A}{C}-\ci{B}{C}=\ci{C}{A}$. Or, for measurements on different parties, $\disc{C}{A}-\disc{C}{B}=\Gamma(C\rangle B)-\Gamma(C\rangle A).$

\section{Regularization}All the relations we have found, although already meaningful in the form above, can be cast in their regularized version, so that they become, in the case of ESM, more consistent from an operational and information-theoretic point of view.  To do so we note that the minimal amount of ebits needed to create $\rho_{AB}$ over all possible LOCC strategies is given by the \emph{entanglement cost} $E_C(A:B) = \lim_{n\rightarrow\infty} \frac{1}{n}E_F(A:B)_{\rho_{AB}^{\otimes n}}$ \cite{regEf}. We can then define the \emph{asymptotic total entanglement consumption of ESM} as the regularized version of Eq. \eqref{cost1}, i.e, as $\Gamma^{\infty}(A\rangle B) := \lim_{n\rightarrow\infty} {\Gamma(A\rangle B)_{\rho_{AB}^{\otimes n}}}/{n}=E_C(A:B)+S(A|B),$ having used that conditional entropy is additive. 
As ESM is itself an asymptotic process, the regularized total cost $\Gamma^{\infty}$ is a quantity better motivated than the unregularized $\Gamma$ from an operational and information-theoretic point of view. It is worth remarking that both $\Gamma$ and $\Gamma^\infty$ are positive, because coherent information is a lower bound on distillable entanglement~\cite{hashinginequality}, and therefore on entanglement cost. By Eq. \eqref{eq:opmeaning} we have that
$\discinf{A}{C}=\Gamma^{\infty}(A\rangle B)$.

\section{Conclusions} We have seen that the QD is intimately related to the tasks of ESM and DC. For a pure tripartite state, the QD reveals what is the entanglement consumption in ESM and in which direction more classical information can be sent through DC. Moreover the asymmetry
of the QD can be given an operational interpretation, since it matches the asymmetry of the tasks to which we have related it, ESM and DC, which are inherently directional.

Finally, a recent paper has unraveled a different
connection between QD and SM \cite{MD}. There, it was observed that the right-hand side of \eqref{disco} can be interpreted
as the difference in quantum communication costs between performing SM with a partially measured
version of $\rho_{AB}$  (first term) and with $\rho_{AB}$ (second term) directly.
Such an interpretation of QD regards a relation between \emph{different} states, one obtained from the other via measurement, while the one presented here refers to just one state (and its purification). On the other hand, since QD can be expressed also as the difference in
mutual information between such two states (see the paragraph after (2)), an approach similar to that of \cite{MD} can lead to interpretations in terms of quantum locking ~\cite{divincenzo_locking_2003,*hayden_randomizing_2004,prep} and correlations erasure \cite{groisman}.

\begin{acknowledgements}
We thank A. Ac\'in, C. B\'eny, J. Calsamiglia, C. Caves, A. Datta, M. Gu, and V. Vedral for valuable comments. This work was supported by the National Research Foundation, the Ministry of Education of Singapore, the Spanish ``Juan de la Cierva" Programme, NSERC, QuantumWorks, Ontario Centres of Excellence, the Royal Society, U.K. EPSRC
and the European Commission.
\end{acknowledgements}


\bibliography{omd.bib}

\begin{thebibliography}{43}%
\makeatletter
\providecommand \@ifxundefined [1]{%
 \@ifx{#1\undefined}
}%
\providecommand \@ifnum [1]{%
 \ifnum #1\expandafter \@firstoftwo
 \else \expandafter \@secondoftwo
 \fi
}%
\providecommand \@ifx [1]{%
 \ifx #1\expandafter \@firstoftwo
 \else \expandafter \@secondoftwo
 \fi
}%
\providecommand \natexlab [1]{#1}%
\providecommand \enquote  [1]{``#1''}%
\providecommand \bibnamefont  [1]{#1}%
\providecommand \bibfnamefont [1]{#1}%
\providecommand \citenamefont [1]{#1}%
\providecommand \href@noop [0]{\@secondoftwo}%
\providecommand \href [0]{\begingroup \@sanitize@url \@href}%
\providecommand \@href[1]{\@@startlink{#1}\@@href}%
\providecommand \@@href[1]{\endgroup#1\@@endlink}%
\providecommand \@sanitize@url [0]{\catcode `\\12\catcode `\$12\catcode
  `\&12\catcode `\#12\catcode `\^12\catcode `\_12\catcode `\%12\relax}%
\providecommand \@@startlink[1]{}%
\providecommand \@@endlink[0]{}%
\providecommand \url  [0]{\begingroup\@sanitize@url \@url }%
\providecommand \@url [1]{\endgroup\@href {#1}{\urlprefix }}%
\providecommand \urlprefix  [0]{URL }%
\providecommand \Eprint [0]{\href }%
\@ifxundefined \urlstyle {%
  \providecommand \doi  [0]{\begingroup \@sanitize@url \@doi}%
  \providecommand \@doi [1]{\endgroup \@@startlink {\doibase
  #1}doi:\discretionary {}{}{}#1\@@endlink }%
}{%
  \providecommand \doi  [0]{doi:\discretionary{}{}{}\begingroup
  \urlstyle{rm}\Url }%
}%
\providecommand \doibase [0]{http://dx.doi.org/}%
\providecommand \Doi [0]{\begingroup \@sanitize@url \@Doi }%
\providecommand \@Doi  [1]{\endgroup\@@startlink{\doibase#1}\@@Doi}%
\providecommand \@@Doi [1]{#1\@@endlink}%
\providecommand \selectlanguage [0]{\@gobble}%
\providecommand \bibinfo  [0]{\@secondoftwo}%
\providecommand \bibfield  [0]{\@secondoftwo}%
\providecommand \translation [1]{[#1]}%
\providecommand \BibitemOpen [0]{}%
\providecommand \bibitemStop [0]{}%
\providecommand \bibitemNoStop [0]{.\EOS\space}%
\providecommand \EOS [0]{\spacefactor3000\relax}%
\providecommand \BibitemShut  [1]{\csname bibitem#1\endcsname}%
\bibitem [{\citenamefont {Horodecki}\ \emph {et~al.}(2009)\citenamefont
  {Horodecki}, \citenamefont {Horodecki}, \citenamefont {Horodecki},\ and\
  \citenamefont {Horodecki}}]{horod}%
  \BibitemOpen
  \bibfield  {author} {\bibinfo {author} {\bibfnamefont {R.}~\bibnamefont
  {Horodecki}}, \bibinfo {author} {\bibfnamefont {P.}~\bibnamefont
  {Horodecki}}, \bibinfo {author} {\bibfnamefont {M.}~\bibnamefont
  {Horodecki}}, \ and\ \bibinfo {author} {\bibfnamefont {K.}~\bibnamefont
  {Horodecki}},\ }\Doi {10.1103/RevModPhys.81.865} {\bibfield  {journal}
  {\bibinfo  {journal} {Rev. Mod. Phys.},\ }\textbf {\bibinfo {volume} {81}},\
  \bibinfo {pages} {865} (\bibinfo {year} {2009})}\BibitemShut {NoStop}%
\bibitem [{\citenamefont {Zurek}(2000)}]{zurek2000}%
  \BibitemOpen
  \bibfield  {author} {\bibinfo {author} {\bibfnamefont {W.~H.}\ \bibnamefont
  {Zurek}},\ }\href@noop {} {\bibfield  {journal} {\bibinfo  {journal} {Annalen
  der Physik (Leipzig)},\ }\textbf {\bibinfo {volume} {9}},\ \bibinfo {pages}
  {855} (\bibinfo {year} {2000})}\BibitemShut {NoStop}%
\bibitem [{\citenamefont {Henderson}\ and\ \citenamefont
  {Vedral}(2001)}]{henderson}%
  \BibitemOpen
  \bibfield  {author} {\bibinfo {author} {\bibfnamefont {L.}~\bibnamefont
  {Henderson}}\ and\ \bibinfo {author} {\bibfnamefont {V.}~\bibnamefont
  {Vedral}},\ }\href@noop {} {\bibfield  {journal} {\bibinfo  {journal} {J.
  Phys. A: Math. Gen.},\ }\textbf {\bibinfo {volume} {34}},\ \bibinfo {pages}
  {6899} (\bibinfo {year} {2001})}\BibitemShut {NoStop}%
\bibitem [{\citenamefont {Ollivier}\ and\ \citenamefont
  {Zurek}(2001)}]{ollivier}%
  \BibitemOpen
  \bibfield  {author} {\bibinfo {author} {\bibfnamefont {H.}~\bibnamefont
  {Ollivier}}\ and\ \bibinfo {author} {\bibfnamefont {W.}~\bibnamefont
  {Zurek}},\ }\href@noop {} {\bibfield  {journal} {\bibinfo  {journal} {Phys.
  Rev. Lett.},\ }\textbf {\bibinfo {volume} {88}},\ \bibinfo {pages} {017901}
  (\bibinfo {year} {2001})}\BibitemShut {NoStop}%
\bibitem [{\citenamefont {Piani}\ \emph {et~al.}(2008)\citenamefont {Piani},
  \citenamefont {Horodecki},\ and\ \citenamefont {Horodecki}}]{piani}%
  \BibitemOpen
  \bibfield  {author} {\bibinfo {author} {\bibfnamefont {M.}~\bibnamefont
  {Piani}}, \bibinfo {author} {\bibfnamefont {P.}~\bibnamefont {Horodecki}}, \
  and\ \bibinfo {author} {\bibfnamefont {R.}~\bibnamefont {Horodecki}},\
  }\href@noop {} {\bibfield  {journal} {\bibinfo  {journal} {Phys. Rev.
  Lett.},\ }\textbf {\bibinfo {volume} {100}},\ \bibinfo {pages} {090502}
  (\bibinfo {year} {2008})}\BibitemShut {NoStop}%
\bibitem [{\citenamefont {Rodriguez-Rosario}\ \emph {et~al.}(2008)\citenamefont
  {Rodriguez-Rosario}, \citenamefont {Modi}, \citenamefont {Kuah},
  \citenamefont {Shaji},\ and\ \citenamefont {Sudarshan}}]{CesarEtal07}%
  \BibitemOpen
  \bibfield  {author} {\bibinfo {author} {\bibfnamefont {C.~A.}\ \bibnamefont
  {Rodriguez-Rosario}}, \bibinfo {author} {\bibfnamefont {K.}~\bibnamefont
  {Modi}}, \bibinfo {author} {\bibfnamefont {A.-M.}\ \bibnamefont {Kuah}},
  \bibinfo {author} {\bibfnamefont {A.}~\bibnamefont {Shaji}}, \ and\ \bibinfo
  {author} {\bibfnamefont {E.~C.~G.}\ \bibnamefont {Sudarshan}},\ }\href@noop
  {} {\bibfield  {journal} {\bibinfo  {journal} {J. Phys. A: Math. Theor.},\
  }\textbf {\bibinfo {volume} {41}},\ \bibinfo {pages} {205301} (\bibinfo
  {year} {2008})}\BibitemShut {NoStop}%
\bibitem [{\citenamefont {Modi}\ \emph {et~al.}(2010)\citenamefont {Modi},
  \citenamefont {Paterek}, \citenamefont {Son}, \citenamefont {Vedral},\ and\
  \citenamefont {Williamson}}]{modietal}%
  \BibitemOpen
  \bibfield  {author} {\bibinfo {author} {\bibfnamefont {K.}~\bibnamefont
  {Modi}}, \bibinfo {author} {\bibfnamefont {T.}~\bibnamefont {Paterek}},
  \bibinfo {author} {\bibfnamefont {W.}~\bibnamefont {Son}}, \bibinfo {author}
  {\bibfnamefont {V.}~\bibnamefont {Vedral}}, \ and\ \bibinfo {author}
  {\bibfnamefont {M.}~\bibnamefont {Williamson}},\ }\href@noop {} {\bibfield
  {journal} {\bibinfo  {journal} {Phys. Rev. Lett.},\ }\textbf {\bibinfo
  {volume} {104}},\ \bibinfo {pages} {080501} (\bibinfo {year}
  {2010})}\BibitemShut {NoStop}%
\bibitem [{\citenamefont {Giorda}\ and\ \citenamefont {Paris}(2010)}]{paris10}%
  \BibitemOpen
  \bibfield  {author} {\bibinfo {author} {\bibfnamefont {P.}~\bibnamefont
  {Giorda}}\ and\ \bibinfo {author} {\bibfnamefont {M.~G.~A.}\ \bibnamefont
  {Paris}},\ }\href@noop {} {\bibfield  {journal} {\bibinfo  {journal} {Phys.
  Rev. Lett.},\ }\textbf {\bibinfo {volume} {105}},\ \bibinfo {pages} {020503}
  (\bibinfo {year} {2010})}\BibitemShut {NoStop}%
\bibitem [{\citenamefont {Adesso}\ and\ \citenamefont
  {Datta}(2010)}]{Adesso10}%
  \BibitemOpen
  \bibfield  {author} {\bibinfo {author} {\bibfnamefont {G.}~\bibnamefont
  {Adesso}}\ and\ \bibinfo {author} {\bibfnamefont {A.}~\bibnamefont {Datta}},\
  }\href@noop {} {\bibfield  {journal} {\bibinfo  {journal} {Phys. Rev.
  Lett.},\ }\textbf {\bibinfo {volume} {105}},\ \bibinfo {pages} {030501}
  (\bibinfo {year} {2010})}\BibitemShut {NoStop}%
\bibitem [{\citenamefont {Daki\'c}\ \emph {et~al.}(2010)\citenamefont
  {Daki\'c}, \citenamefont {Vedral},\ and\ \citenamefont {\v{C}.
  Brukner}}]{dvb10}%
  \BibitemOpen
  \bibfield  {author} {\bibinfo {author} {\bibfnamefont {B.}~\bibnamefont
  {Daki\'c}}, \bibinfo {author} {\bibfnamefont {V.}~\bibnamefont {Vedral}}, \
  and\ \bibinfo {author} {\bibnamefont {\v{C}. Brukner}},\ }\href@noop {}
  {\bibfield  {journal} {\bibinfo  {journal} {arXiv:1004.0190}} (\bibinfo
  {year} {2010})}\BibitemShut {NoStop}%
\bibitem [{\citenamefont {Ferraro}\ \emph {et~al.}(2010)\citenamefont
  {Ferraro}, \citenamefont {Aolita}, \citenamefont {Cavalcanti}, \citenamefont
  {Cucchietti},\ and\ \citenamefont {Ac\'in}}]{ferraro}%
  \BibitemOpen
  \bibfield  {author} {\bibinfo {author} {\bibfnamefont {A.}~\bibnamefont
  {Ferraro}}, \bibinfo {author} {\bibfnamefont {L.}~\bibnamefont {Aolita}},
  \bibinfo {author} {\bibfnamefont {D.}~\bibnamefont {Cavalcanti}}, \bibinfo
  {author} {\bibfnamefont {F.~M.}\ \bibnamefont {Cucchietti}}, \ and\ \bibinfo
  {author} {\bibfnamefont {A.}~\bibnamefont {Ac\'in}},\ }\href@noop {}
  {\bibfield  {journal} {\bibinfo  {journal} {Phys. Rev. A},\ }\textbf
  {\bibinfo {volume} {81}},\ \bibinfo {pages} {052318} (\bibinfo {year}
  {2010})}\BibitemShut {NoStop}%
\bibitem [{\citenamefont {Werlang}\ \emph {et~al.}(2009)\citenamefont
  {Werlang}, \citenamefont {Souza}, \citenamefont {Fanchini},\ and\
  \citenamefont {Villas-Boas}}]{werlangPRA}%
  \BibitemOpen
  \bibfield  {author} {\bibinfo {author} {\bibfnamefont {T.}~\bibnamefont
  {Werlang}}, \bibinfo {author} {\bibfnamefont {S.}~\bibnamefont {Souza}},
  \bibinfo {author} {\bibfnamefont {F.~F.}\ \bibnamefont {Fanchini}}, \ and\
  \bibinfo {author} {\bibfnamefont {C.~J.}\ \bibnamefont {Villas-Boas}},\
  }\href@noop {} {\bibfield  {journal} {\bibinfo  {journal} {Phys. Rev. A},\
  }\textbf {\bibinfo {volume} {80}},\ \bibinfo {pages} {024103} (\bibinfo
  {year} {2009})}\BibitemShut {NoStop}%
\bibitem [{\citenamefont {Maziero}\ \emph {et~al.}(2009)\citenamefont
  {Maziero}, \citenamefont {Celeri}, \citenamefont {Serra},\ and\ \citenamefont
  {Vedral}}]{Maziero09}%
  \BibitemOpen
  \bibfield  {author} {\bibinfo {author} {\bibfnamefont {J.}~\bibnamefont
  {Maziero}}, \bibinfo {author} {\bibfnamefont {L.~C.}\ \bibnamefont {Celeri}},
  \bibinfo {author} {\bibfnamefont {R.~M.}\ \bibnamefont {Serra}}, \ and\
  \bibinfo {author} {\bibfnamefont {V.}~\bibnamefont {Vedral}},\ }\href@noop {}
  {\bibfield  {journal} {\bibinfo  {journal} {Phys. Rev. A},\ }\textbf
  {\bibinfo {volume} {80}},\ \bibinfo {pages} {044102} (\bibinfo {year}
  {2009})}\BibitemShut {NoStop}%
\bibitem [{\citenamefont {Mazzola}\ \emph {et~al.}(2010)\citenamefont
  {Mazzola}, \citenamefont {Piilo},\ and\ \citenamefont
  {Maniscalco}}]{mazzola}%
  \BibitemOpen
  \bibfield  {author} {\bibinfo {author} {\bibfnamefont {L.}~\bibnamefont
  {Mazzola}}, \bibinfo {author} {\bibfnamefont {J.}~\bibnamefont {Piilo}}, \
  and\ \bibinfo {author} {\bibfnamefont {S.}~\bibnamefont {Maniscalco}},\
  }\href@noop {} {\bibfield  {journal} {\bibinfo  {journal} {Phys. Rev.
  Lett.},\ }\textbf {\bibinfo {volume} {104}},\ \bibinfo {pages} {200401}
  (\bibinfo {year} {2010})}\BibitemShut {NoStop}%
\bibitem [{\citenamefont {Fanchini}\ \emph
  {et~al.}(2010){\natexlab{a}}\citenamefont {Fanchini}, \citenamefont
  {Werlang}, \citenamefont {Brasil}, \citenamefont {Arruda},\ and\
  \citenamefont {Caldeira}}]{fanchini}%
  \BibitemOpen
  \bibfield  {author} {\bibinfo {author} {\bibfnamefont {F.~F.}\ \bibnamefont
  {Fanchini}}, \bibinfo {author} {\bibfnamefont {T.}~\bibnamefont {Werlang}},
  \bibinfo {author} {\bibfnamefont {C.~A.}\ \bibnamefont {Brasil}}, \bibinfo
  {author} {\bibfnamefont {L.~G.~E.}\ \bibnamefont {Arruda}}, \ and\ \bibinfo
  {author} {\bibfnamefont {A.~O.}\ \bibnamefont {Caldeira}},\ }\href@noop {}
  {\bibfield  {journal} {\bibinfo  {journal} {Phys. Rev. A},\ }\textbf
  {\bibinfo {volume} {81}},\ \bibinfo {pages} {052107} (\bibinfo {year}
  {2010}{\natexlab{a}})}\BibitemShut {NoStop}%
\bibitem [{\citenamefont {Xu}\ \emph {et~al.}(2010)\citenamefont {Xu} \emph
  {et~al.}}]{jsxu}%
  \BibitemOpen
  \bibfield  {author} {\bibinfo {author} {\bibfnamefont {J.-S.}\ \bibnamefont
  {Xu}} \emph {et~al.},\ }\href@noop {} {\bibfield  {journal} {\bibinfo
  {journal} {Nat. Comm.},\ }\textbf {\bibinfo {volume} {1}},\ \bibinfo {pages}
  {7} (\bibinfo {year} {2010})}\BibitemShut {NoStop}%
\bibitem [{\citenamefont {Datta}\ \emph {et~al.}(2008)\citenamefont {Datta},
  \citenamefont {Shaji},\ and\ \citenamefont {Caves}}]{dattashaji}%
  \BibitemOpen
  \bibfield  {author} {\bibinfo {author} {\bibfnamefont {A.}~\bibnamefont
  {Datta}}, \bibinfo {author} {\bibfnamefont {A.}~\bibnamefont {Shaji}}, \ and\
  \bibinfo {author} {\bibfnamefont {C.}~\bibnamefont {Caves}},\ }\href@noop {}
  {\bibfield  {journal} {\bibinfo  {journal} {Phys. Rev. Lett.},\ }\textbf
  {\bibinfo {volume} {100}},\ \bibinfo {pages} {050502} (\bibinfo {year}
  {2008})}\BibitemShut {NoStop}%
\bibitem [{\citenamefont {Lanyon}\ \emph {et~al.}(2008)\citenamefont {Lanyon},
  \citenamefont {Barbieri}, \citenamefont {Almeida},\ and\ \citenamefont
  {White}}]{Lanyon}%
  \BibitemOpen
  \bibfield  {author} {\bibinfo {author} {\bibfnamefont {B.~P.}\ \bibnamefont
  {Lanyon}}, \bibinfo {author} {\bibfnamefont {M.}~\bibnamefont {Barbieri}},
  \bibinfo {author} {\bibfnamefont {M.~P.}\ \bibnamefont {Almeida}}, \ and\
  \bibinfo {author} {\bibfnamefont {A.~G.}\ \bibnamefont {White}},\ }\href@noop
  {} {\bibfield  {journal} {\bibinfo  {journal} {Phys. Rev. Lett.},\ }\textbf
  {\bibinfo {volume} {101}},\ \bibinfo {pages} {200501} (\bibinfo {year}
  {2008})}\BibitemShut {NoStop}%
\bibitem [{\citenamefont {Sarandy}(2009)}]{sarandy}%
  \BibitemOpen
  \bibfield  {author} {\bibinfo {author} {\bibfnamefont {M.}~\bibnamefont
  {Sarandy}},\ }\href@noop {} {\bibfield  {journal} {\bibinfo  {journal} {Phys.
  Rev. A},\ }\textbf {\bibinfo {volume} {80}},\ \bibinfo {pages} {022108}
  (\bibinfo {year} {2009})}\BibitemShut {NoStop}%
\bibitem [{\citenamefont {Werlang}\ \emph {et~al.}(2010)\citenamefont
  {Werlang}, \citenamefont {C.Trippe}, \citenamefont {Ribeiro},\ and\
  \citenamefont {Rigolin}}]{werlangQPT}%
  \BibitemOpen
  \bibfield  {author} {\bibinfo {author} {\bibfnamefont {T.}~\bibnamefont
  {Werlang}}, \bibinfo {author} {\bibnamefont {C.Trippe}}, \bibinfo {author}
  {\bibfnamefont {G.}~\bibnamefont {Ribeiro}}, \ and\ \bibinfo {author}
  {\bibfnamefont {G.}~\bibnamefont {Rigolin}},\ }\href@noop {} {\bibfield
  {journal} {\bibinfo  {journal} {Phys. Rev. Lett.},\ }\textbf {\bibinfo
  {volume} {105}},\ \bibinfo {pages} {095702} (\bibinfo {year}
  {2010})}\BibitemShut {NoStop}%
\bibitem [{\citenamefont {Zurek}(2003)}]{ZurekDemon}%
  \BibitemOpen
  \bibfield  {author} {\bibinfo {author} {\bibfnamefont {W.~H.}\ \bibnamefont
  {Zurek}},\ }\href@noop {} {\bibfield  {journal} {\bibinfo  {journal} {Phys.
  Rev. A},\ }\textbf {\bibinfo {volume} {67}},\ \bibinfo {pages} {012320}
  (\bibinfo {year} {2003})}\BibitemShut {NoStop}%
\bibitem [{\citenamefont {Brodutch}\ and\ \citenamefont {Terno}(2010)}]{Terno}%
  \BibitemOpen
  \bibfield  {author} {\bibinfo {author} {\bibfnamefont {A.}~\bibnamefont
  {Brodutch}}\ and\ \bibinfo {author} {\bibfnamefont {D.~R.}\ \bibnamefont
  {Terno}},\ }\href@noop {} {\bibfield  {journal} {\bibinfo  {journal} {Phys.
  Rev. A},\ }\textbf {\bibinfo {volume} {81}},\ \bibinfo {pages} {062103}
  (\bibinfo {year} {2010})}\BibitemShut {NoStop}%
\bibitem [{\citenamefont {Horodecki}\ \emph {et~al.}(2005)\citenamefont
  {Horodecki}, \citenamefont {Oppenheim},\ and\ \citenamefont
  {Winter}}]{merging}%
  \BibitemOpen
  \bibfield  {author} {\bibinfo {author} {\bibfnamefont {M.}~\bibnamefont
  {Horodecki}}, \bibinfo {author} {\bibfnamefont {J.}~\bibnamefont
  {Oppenheim}}, \ and\ \bibinfo {author} {\bibfnamefont {A.}~\bibnamefont
  {Winter}},\ }\href@noop {} {\bibfield  {journal} {\bibinfo  {journal}
  {Nature},\ }\textbf {\bibinfo {volume} {436}} (\bibinfo {year}
  {2005})}\BibitemShut {NoStop}%
\bibitem [{\citenamefont {Bennett}\ and\ \citenamefont
  {Wiesner}(1992)}]{bennett92}%
  \BibitemOpen
  \bibfield  {author} {\bibinfo {author} {\bibfnamefont {C.~H.}\ \bibnamefont
  {Bennett}}\ and\ \bibinfo {author} {\bibfnamefont {S.~J.}\ \bibnamefont
  {Wiesner}},\ }\href@noop {} {\bibfield  {journal} {\bibinfo  {journal} {Phys.
  Rev. Lett.},\ }\textbf {\bibinfo {volume} {69}},\ \bibinfo {pages} {2881}
  (\bibinfo {year} {1992})}\BibitemShut {NoStop}%
\bibitem [{\citenamefont {Nielsen}\ and\ \citenamefont
  {Chuang}(2000)}]{nielsen}%
  \BibitemOpen
  \bibfield  {author} {\bibinfo {author} {\bibfnamefont {M.~A.}\ \bibnamefont
  {Nielsen}}\ and\ \bibinfo {author} {\bibfnamefont {I.~L.}\ \bibnamefont
  {Chuang}},\ }\href@noop {} {\emph {\bibinfo {title} {Quantum Computation and
  Quantum Information}}}\ (\bibinfo  {publisher} {Cambridge University Press},\
  \bibinfo {address} {Cambridge, UK},\ \bibinfo {year} {2000})\BibitemShut
  {NoStop}%
\bibitem [{\citenamefont {Slepian}\ and\ \citenamefont
  {Wolf}(1971)}]{slepianwolf}%
  \BibitemOpen
  \bibfield  {author} {\bibinfo {author} {\bibfnamefont {D.}~\bibnamefont
  {Slepian}}\ and\ \bibinfo {author} {\bibfnamefont {J.~K.}\ \bibnamefont
  {Wolf}},\ }\href@noop {} {\bibfield  {journal} {\bibinfo  {journal} {IEEE
  Trans. Inf. Theory},\ }\textbf {\bibinfo {volume} {19}},\ \bibinfo {pages}
  {461} (\bibinfo {year} {1971})}\BibitemShut {NoStop}%
\bibitem [{\citenamefont {Schumacher}\ and\ \citenamefont
  {Nielsen}(1996)}]{Schumacher}%
  \BibitemOpen
  \bibfield  {author} {\bibinfo {author} {\bibfnamefont {B.}~\bibnamefont
  {Schumacher}}\ and\ \bibinfo {author} {\bibfnamefont {M.~A.}\ \bibnamefont
  {Nielsen}},\ }\href@noop {} {\bibfield  {journal} {\bibinfo  {journal} {Phys.
  Rev. A},\ }\textbf {\bibinfo {volume} {54}},\ \bibinfo {pages} {2629}
  (\bibinfo {year} {1996})}\BibitemShut {NoStop}%
\bibitem [{Note1()}]{Note1}%
  \BibitemOpen
  \bibinfo {note} {In \cite {ollivier} the generalized measurement were
  actually restricted to complete von Neumann measurements.}\BibitemShut
  {Stop}%
\bibitem [{\citenamefont {Groisman}\ \emph {et~al.}(2005)\citenamefont
  {Groisman}, \citenamefont {Popescu},\ and\ \citenamefont
  {Winter}}]{groisman}%
  \BibitemOpen
  \bibfield  {author} {\bibinfo {author} {\bibfnamefont {B.}~\bibnamefont
  {Groisman}}, \bibinfo {author} {\bibfnamefont {S.}~\bibnamefont {Popescu}}, \
  and\ \bibinfo {author} {\bibfnamefont {A.}~\bibnamefont {Winter}},\
  }\href@noop {} {\bibfield  {journal} {\bibinfo  {journal} {Phys. Rev. A},\
  }\textbf {\bibinfo {volume} {72}},\ \bibinfo {pages} {032317} (\bibinfo
  {year} {2005})}\BibitemShut {NoStop}%
\bibitem [{\citenamefont {Bennett}\ \emph
  {et~al.}(1996){\natexlab{a}}\citenamefont {Bennett}, \citenamefont
  {DiVincenzo}, \citenamefont {Smolin},\ and\ \citenamefont
  {Wootters}}]{bennett}%
  \BibitemOpen
  \bibfield  {author} {\bibinfo {author} {\bibfnamefont {C.~H.}\ \bibnamefont
  {Bennett}}, \bibinfo {author} {\bibfnamefont {D.~P.}\ \bibnamefont
  {DiVincenzo}}, \bibinfo {author} {\bibfnamefont {J.~A.}\ \bibnamefont
  {Smolin}}, \ and\ \bibinfo {author} {\bibfnamefont {W.~K.}\ \bibnamefont
  {Wootters}},\ }\href@noop {} {\bibfield  {journal} {\bibinfo  {journal}
  {Phys. Rev. A},\ }\textbf {\bibinfo {volume} {54}},\ \bibinfo {pages} {3824}
  (\bibinfo {year} {1996}{\natexlab{a}})}\BibitemShut {NoStop}%
\bibitem [{\citenamefont {Horodecki}\ \emph {et~al.}(2001)\citenamefont
  {Horodecki}, \citenamefont {Horodecki}, \citenamefont {Horodecki},
  \citenamefont {Leung},\ and\ \citenamefont {Terhal}}]{DC}%
  \BibitemOpen
  \bibfield  {author} {\bibinfo {author} {\bibfnamefont {M.}~\bibnamefont
  {Horodecki}}, \bibinfo {author} {\bibfnamefont {P.}~\bibnamefont
  {Horodecki}}, \bibinfo {author} {\bibfnamefont {R.}~\bibnamefont
  {Horodecki}}, \bibinfo {author} {\bibfnamefont {D.}~\bibnamefont {Leung}}, \
  and\ \bibinfo {author} {\bibfnamefont {B.}~\bibnamefont {Terhal}},\
  }\href@noop {} {\bibfield  {journal} {\bibinfo  {journal} {Quantum Inf.
  Comput.},\ }\textbf {\bibinfo {volume} {1}},\ \bibinfo {pages} {70} (\bibinfo
  {year} {2001})}\BibitemShut {NoStop}%
\bibitem [{\citenamefont {Winter}(2002)}]{winterdense}%
  \BibitemOpen
  \bibfield  {author} {\bibinfo {author} {\bibfnamefont {A.}~\bibnamefont
  {Winter}},\ }\href@noop {} {\bibfield  {journal} {\bibinfo  {journal} {J.
  Math. Phys.},\ }\textbf {\bibinfo {volume} {43}},\ \bibinfo {pages} {4341}
  (\bibinfo {year} {2002})}\BibitemShut {NoStop}%
\bibitem [{\citenamefont {Bru{\ss}}\ \emph {et~al.}(2004)\citenamefont
  {Bru{\ss}} \emph {et~al.}}]{Bruss}%
  \BibitemOpen
  \bibfield  {author} {\bibinfo {author} {\bibfnamefont {D.}~\bibnamefont
  {Bru{\ss}}} \emph {et~al.},\ }\href@noop {} {\bibfield  {journal} {\bibinfo
  {journal} {Phys. Rev. Lett.},\ }\textbf {\bibinfo {volume} {93}},\ \bibinfo
  {pages} {210501} (\bibinfo {year} {2004})}\BibitemShut {NoStop}%
\bibitem [{\citenamefont {Horodecki}\ and\ \citenamefont
  {Piani}(2007)}]{HoroPiani}%
  \BibitemOpen
  \bibfield  {author} {\bibinfo {author} {\bibfnamefont {M.}~\bibnamefont
  {Horodecki}}\ and\ \bibinfo {author} {\bibfnamefont {M.}~\bibnamefont
  {Piani}},\ }\href@noop {} {\bibfield  {journal} {\bibinfo  {journal}
  {arXiv:quant-ph/0701134}} (\bibinfo {year} {2007})}\BibitemShut {NoStop}%
\bibitem [{\citenamefont {Bennett}\ \emph
  {et~al.}(1996){\natexlab{b}}\citenamefont {Bennett}, \citenamefont
  {DiVincenzo}, \citenamefont {Smolin},\ and\ \citenamefont
  {Wootters}}]{regEf}%
  \BibitemOpen
  \bibfield  {author} {\bibinfo {author} {\bibfnamefont {C.~H.}\ \bibnamefont
  {Bennett}}, \bibinfo {author} {\bibfnamefont {D.~P.}\ \bibnamefont
  {DiVincenzo}}, \bibinfo {author} {\bibfnamefont {J.~A.}\ \bibnamefont
  {Smolin}}, \ and\ \bibinfo {author} {\bibfnamefont {W.~K.}\ \bibnamefont
  {Wootters}},\ }\href@noop {} {\bibfield  {journal} {\bibinfo  {journal}
  {Phys. Rev. A},\ }\textbf {\bibinfo {volume} {54}},\ \bibinfo {pages} {3824}
  (\bibinfo {year} {1996}{\natexlab{b}})}\BibitemShut {NoStop}%
\bibitem [{\citenamefont {Devetak}\ and\ \citenamefont
  {Winter}(2005)}]{hashinginequality}%
  \BibitemOpen
  \bibfield  {author} {\bibinfo {author} {\bibfnamefont {I.}~\bibnamefont
  {Devetak}}\ and\ \bibinfo {author} {\bibfnamefont {A.}~\bibnamefont
  {Winter}},\ }\href@noop {} {\bibfield  {journal} {\bibinfo  {journal} {Proc.
  R. Soc. Lond. A},\ }\textbf {\bibinfo {volume} {461}},\ \bibinfo {pages}
  {207} (\bibinfo {year} {2005})}\BibitemShut {NoStop}%
\bibitem [{\citenamefont {Madhok}\ and\ \citenamefont {Datta}(2010)}]{MD}%
  \BibitemOpen
  \bibfield  {author} {\bibinfo {author} {\bibfnamefont {V.}~\bibnamefont
  {Madhok}}\ and\ \bibinfo {author} {\bibfnamefont {A.}~\bibnamefont {Datta}},\
  }\href@noop {} {\bibfield  {journal} {\bibinfo  {journal} {arXiv:1008.4135}}
  (\bibinfo {year} {2010})}\BibitemShut {NoStop}%
\bibitem [{\citenamefont {DiVincenzo}\ \emph {et~al.}(2003)\citenamefont
  {DiVincenzo}, \citenamefont {Horodecki}, \citenamefont {Leung}, \citenamefont
  {Smolin},\ and\ \citenamefont {Terhal}}]{divincenzo_locking_2003}%
  \BibitemOpen
  \bibfield  {author} {\bibinfo {author} {\bibfnamefont {D.~P.}\ \bibnamefont
  {DiVincenzo}}, \bibinfo {author} {\bibfnamefont {M.}~\bibnamefont
  {Horodecki}}, \bibinfo {author} {\bibfnamefont {D.~W.}\ \bibnamefont
  {Leung}}, \bibinfo {author} {\bibfnamefont {J.~A.}\ \bibnamefont {Smolin}}, \
  and\ \bibinfo {author} {\bibfnamefont {B.~M.}\ \bibnamefont {Terhal}},\
  }\href@noop {} {\bibfield  {journal} {\bibinfo  {journal} {Phys. Rev.
  Lett.},\ }\textbf {\bibinfo {volume} {92}},\ \bibinfo {pages} {067902}
  (\bibinfo {year} {2003})}\BibitemShut {NoStop}%
\bibitem [{\citenamefont {Hayden}\ \emph {et~al.}(2004)\citenamefont {Hayden},
  \citenamefont {Leung}, \citenamefont {Shor},\ and\ \citenamefont
  {Winter}}]{hayden_randomizing_2004}%
  \BibitemOpen
  \bibfield  {author} {\bibinfo {author} {\bibfnamefont {P.}~\bibnamefont
  {Hayden}}, \bibinfo {author} {\bibfnamefont {D.}~\bibnamefont {Leung}},
  \bibinfo {author} {\bibfnamefont {P.~W.}\ \bibnamefont {Shor}}, \ and\
  \bibinfo {author} {\bibfnamefont {A.}~\bibnamefont {Winter}},\ }\href@noop {}
  {\bibfield  {journal} {\bibinfo  {journal} {Comm. Math. Phys.},\ }\textbf
  {\bibinfo {volume} {250}},\ \bibinfo {pages} {371} (\bibinfo {year}
  {2004})}\BibitemShut {NoStop}%
\bibitem [{\citenamefont {Boixo}\ \emph {et~al.}(2010)\citenamefont {Boixo}
  \emph {et~al.}}]{prep}%
  \BibitemOpen
  \bibfield  {author} {\bibinfo {author} {\bibfnamefont {S.}~\bibnamefont
  {Boixo}} \emph {et~al.},\ }\href@noop {} {\bibfield  {journal} {\bibinfo
  {journal} {in preparation}} (\bibinfo {year} {2010})}\BibitemShut {NoStop}%
\bibitem [{\citenamefont {Koashi}\ and\ \citenamefont
  {Winter}(2004)}]{KoashiWinter}%
  \BibitemOpen
  \bibfield  {author} {\bibinfo {author} {\bibfnamefont {M.}~\bibnamefont
  {Koashi}}\ and\ \bibinfo {author} {\bibfnamefont {A.}~\bibnamefont
  {Winter}},\ }\href@noop {} {\bibfield  {journal} {\bibinfo  {journal} {Phys.
  Rev. A},\ }\textbf {\bibinfo {volume} {69}},\ \bibinfo {pages} {022309}
  (\bibinfo {year} {2004})}\BibitemShut {NoStop}%
\bibitem [{\citenamefont {Fanchini}\ \emph
  {et~al.}(2010){\natexlab{b}}\citenamefont {Fanchini}, \citenamefont
  {Cornelio}, \citenamefont {de~Oliveira},\ and\ \citenamefont
  {Caldeira}}]{fanchini1}%
  \BibitemOpen
  \bibfield  {author} {\bibinfo {author} {\bibfnamefont {F.~F.}\ \bibnamefont
  {Fanchini}}, \bibinfo {author} {\bibfnamefont {M.~F.}\ \bibnamefont
  {Cornelio}}, \bibinfo {author} {\bibfnamefont {M.~C.}\ \bibnamefont
  {de~Oliveira}}, \ and\ \bibinfo {author} {\bibfnamefont {A.~O.}\ \bibnamefont
  {Caldeira}},\ }\href@noop {} {\bibfield  {journal} {\bibinfo  {journal}
  {arXiv:1006.2460v1.}} (\bibinfo {year} {2010}{\natexlab{b}})}\BibitemShut
  {NoStop}%
\bibitem [{\citenamefont {Terhal}\ \emph {et~al.}(2002)\citenamefont {Terhal},
  \citenamefont {Horodecki}, \citenamefont {DiVincenzo},\ and\ \citenamefont
  {Leung}}]{IBMHor2002}%
  \BibitemOpen
  \bibfield  {author} {\bibinfo {author} {\bibfnamefont {B.~M.}\ \bibnamefont
  {Terhal}}, \bibinfo {author} {\bibfnamefont {M.}~\bibnamefont {Horodecki}},
  \bibinfo {author} {\bibfnamefont {D.~P.}\ \bibnamefont {DiVincenzo}}, \ and\
  \bibinfo {author} {\bibfnamefont {D.}~\bibnamefont {Leung}},\ }\href@noop {}
  {\bibfield  {journal} {\bibinfo  {journal} {J. Math Phys.},\ }\textbf
  {\bibinfo {volume} {43}},\ \bibinfo {pages} {4286} (\bibinfo {year}
  {2002})}\BibitemShut {NoStop}%
\end{thebibliography}%

\section{Appendix 1: Proof of Eq. (4)}
We start by recalling the Koashi-Winter monogamy relation \cite{KoashiWinter} for quantum correlations within a pure tripartite state $\psi_{ABC}$:
\be\label{eq:KW}
S(B)=E_F(A:B)+I(B:C_c).
\ee
This, together with the definition of $I(B:C_c)$, implies that 
\be
E_F(A:B)=S(B|C_c)=S(A|C_c),
\ee
 which we can substitute in the definition of $\disc{A}{C}$ to get~\cite{fanchini1}
 \be \disc{A}{C}=E_F(A:B)-S(A|C).
 \ee 
Now, note that $S(A|C)=S(AC)-S(C)$ and, since $\psi_{ABC}$ is a pure state, we have $S(AC)=S(B)$ and $S(C)=S(AB)$. 
Hence $S(A|C)=S(B)-S(AB)=-S(A|B)$, so that
\be
\disc{A}{C}=E_F(A:B)+S(A|B)=\tec{A}{B}.
\ee

\section{Appendix 2: Proof of Eq. (7)}
A monogamy equality similar to Eq. \eqref{eq:KW} with regards to DC was given in \cite{HoroPiani}:
\be
\label{eq:HP}
S(A)=E_P(A:C)+\Ddc{B}{A},
\ee
where $E_P$ is the \emph{entanglement of purification}, defined as~\cite{IBMHor2002} $E_P(A:C):=\min_{\psi_{AA'CC'}} S\big( \trace_{CC'} (\psi_{AA'CC'}) \big)$, with the minimum taken over all pure states $\psi_{AA'CC'}$ such that $\trace_{A'C'}(\psi_{AA'CC'})=\rho_{AC}$. Using the fact that for a tripartite pure state $\ci{A}{C}=S(C)-S(B)$, and expressing $S(B)$ according to \eqref{eq:HP}, from \eqref{eq:opmeaning} one obtains $\disc{A}{C} = S(C) - \Ddc{C}{B} - \big( E_P(A:B) - \eof{A}{B} \big)$. 
Applying this equivalence twice one gets
\be
\disc{A}{C}-\disc{B}{C}=\Ddc{C}{A}-\Ddc{C}{B}.
\ee

\end{document}